\documentclass{evn2004}
\usepackage{txfonts}
\usepackage{graphicx}
\begin{document}
\setcounter{page}{327}

   \title{Multi-wavelength differential astrometry of the S5 polar 
   cap sample
          }
\titlerunning{S5 polar cap sample astrometry}

   \author{J.C. Guirado\inst{1}, J.M. Marcaide\inst{1},
          E. Ros\inst{2}, M.A. P\'erez-Torres\inst{3}, 
          \and
          I. Mart\'{\i}-Vidal\inst{1}
          }

\authorrunning{Guirado et al.}

\institute{Departamento de Astronom\'{\i}a y Astrof\'{\i}sica, Universidad de Valencia, 
E-46100 Burjassot, Valencia, Spain
\and Max-Planck Institut f\"ur Radioastronomie, Auf dem H\"ugel 69, D-53121 Bonn, Germany 
\and Instituto de Astrof\'{\i}sica de Andaluc\'{\i}a (CSIC), Apdo. Correos 3004, E-18080 Granada, Spain } 

\abstract{We report on the status of our S5 polar cap astrometry program. Since 1997 we have observed 
all the 13 radio sources of the complete 
S5 polar cap sample at the wavelengths of 3.6\,cm, 2\,cm and 
0.7\,mm. 
Images of the radio sources at 3.6 and 2\,cm have already been published 
reporting morphological changes. Preliminary astrometric analyses have been carried out at three 
frequencies with 
precisions in the relative position determination ranging from 80 to 
20$\mu$as. We report also on the combination of our phase-delay global 
astrometry results with the $\mu$as-precise optical astrometry that 
will be provided by future space-based instruments. }

   \maketitle
%

\section{Introduction}

\noindent
We are conducting a multi-wavelength 
astrometric study of the ``complete S5 polar cap sample", consisting of 
thirteen radio sources from the 
S5 survey (K\"uhr et al. 1981, Eckart et al. 1986) defined by the 
following criteria: (i) $\delta \ge 70\degr$, 
(ii) $\mid b_{II}\mid \ge 10\degr$, 
(iii) $S_{5\,\mathrm{GHz}}\ge 1\,Jy$ at the epoch of the survey, and 
(iv) $\alpha_{2.7, 5\,\mathrm{GHz}}\ge$ $-$0.5 ($S \sim \nu^{+\alpha}$). 
All sources in this 
sample have large flux densities, well defined ICRF (International 
Celestial Reference Frame; Ma et al. 1998) positions, and 
relative separations less than about $15^\circ$, which 
guarantees astrometric precisions better than 0.1\,mas via the 
application of 
phase-delay differential astrometry techniques. The goal of 
this program is to determine the absolute kinematics of all 
the sources using maps properly registered through different epochs.
Given the variety of source structures in the sample and its 
completeness, our program will result in a 
definitive check of the standard jet model 
(Blandford \& K\"onigl 1979). \\

\noindent
We have observed all 13 members of the sample at 
$\lambda$3.6\,cm, 
at $\lambda$2\,cm and at $\lambda$7\,mm (see Table 1) with 
the VLBA, using a multiple triangulation approach. Data were 
correlated at the VLBA Array Operations Center. In each of the 24-hour 
observations at each wavelength, each source has been tracked over an average 
time of 5\,hr (with a total integration time of 2\,hr), enough to produce 
high-quality hybrid maps. In this contribution we report briefly some 
astrometric results from the epochs already analyzed.

\begin{table}
\begin{center}
\caption[b]{S5 polar-cap sample observations}
\begin{tabular}{|l|ccc|} \hline 

{\bf Epoch} & \multicolumn{3}{c|}{\bf Wavelength} \\ 

        & $\lambda$3.6\,cm & $\lambda$2\,cm & $\lambda$7\,mm \\ \hline
1997.93 &  $\surd$         &               &                \\ 
1999.41 &  $\surd$                &               &                \\ 
1999.57 &                  &  $\surd$             &                \\ 
2000.46 &                  &  $\surd$             &                \\ 
2001.04 &                  &  $\surd$             &    $\surd$            \\ 
2001.09 &  $\surd$                &               &                \\ 
2001.71 &                  &               &   $\surd$             \\ 
2004.62 &                  &  $\surd$             &    $\surd$     \\ \hline 

\end{tabular}
\end{center}
\end{table}

\begin{figure*}[htb]
\vspace{9cm}
\includegraphics{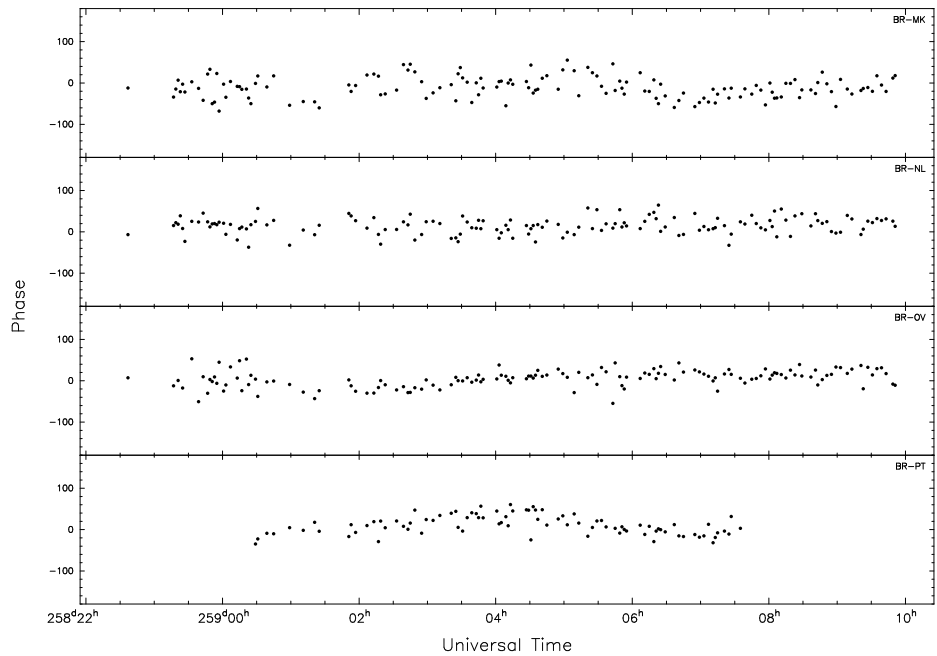}
\caption[] 
{\small Residual $\lambda$7\,mm phases for 2007+777 referenced to those 
of 1928+738 for a representative set of baselines at epoch 2001.71. 
Notice the absence of systematic effects. The average root-mean-square noise 
of the residuals is $\sim$3\,ps (one phase-cycle at $\lambda$7\,mm corresponds 
to 23\,ps of phase-delay).}

\label{fig:phas-conn}
\end{figure*}

\section{Results}

\noindent
The maps of all 13 radio sources for the first two epochs 
at $\lambda$3.6\,cm and $\lambda$2\,cm 
have already been published (Ros et al. 2001; P\'erez-Torres et al. 2004). At 
these wavelengths, these authors found that 
most of the sources of the sample have 
one-sided jet structures, but also found the existence of intriguing compact 
structures (i.e., 0615+820). Combining maps of different epochs, they have 
analyzed 
and modeled the strong morphological changes found in some sources 
(i.e., 0016+731, 0836+719, 1928+738, and 2007+777, amongst others). Combining 
maps at both frequencies, and using spectral index estimates, they have 
determined that for most sources the brightest feature can be identified 
with the core. These findings are essential 
to properly define a suitable astrometric reference point on the structure of
each source, and will be necessary for a meaningful interpretation of the 
astrometric results. \\

\noindent
Preliminary phase-delay astrometric work has been carried out at 
all frequencies, involving all data jointly and using bootstrapping 
techniques (see Ros et al. 1998). The 
root-mean-square noise of the differenced phase-delay residuals ranges from $\sim$30\,ps 
at $\lambda$3.6\,cm (Ros et al. 1998) to $\sim$3\,ps at  $\lambda$7\,mm 
(similar to that obtained by 
Guirado et al. 2000; see Fig. 1).
If we take into account all systematic errors, 
the average precision of the relative position determination of all 
43 radio source
pairs in the sample is $\sim$80$\mu$as at $\lambda$3.6\,cm, 
$\sim$50$\mu$as at 2\,cm and $\sim$20$\mu$as at 7\,mm. With a 
time span of more than seven years, our program will set bounds of 
5-10 $\mu$as/yr to the proper motion of the core of the S5 sources.
This result will show that the study of the absolute 
kinematics of a complete sample of 
sources can be done, almost routinely, using the phase-delay 
observable.\\

\noindent
Our program opens the possibility to extend phase-delay astrometry to 
large portions of the sky, if not the entire sky, acting as a more precise 
complement of the ICRF realizations. In a sense, there is a need for 
such $\mu$as-precise global astrometry; a variety of AGN phenomena 
are expected to be observed by space-based astrometric 
missions, such as the Space Interferometry Mission, SIM (Unwin \& Shao 2000), 
which will yield $\mu$as-precise positions, 
and proper motion of a few $\mu$as/yr (Unwin et al. 2002), at levels comparable
with those obtained in our project. The well-known structures of the 
S5 sources and the precise monitoring of the absolute kinematics, when combined 
with SIM optical data, will be essential to answer fundamental questions 
regarding stability of the radio/optical reference frame tie, or the 
location of the optical emission. The combination of both high-precision radio 
and optical positions will show that astrometry is a useful tool 
to understand the nature of AGNs.

\begin{acknowledgements}
This work has been partially supported by Grant AYA2001-2147-C02-02 of 
the Spanish DGICYT. MAPT acknowledges the support of a Ram\'on y 
Cajal Fellowship from the Spanish MCyT. IMV acknowledges financial 
support of MCyT, Spain, grant BES-2004-5089. The National Radio Astronomy Observatory is 
operated by Associated Universities, Inc., under cooperative agreement 
with the National Science Foundation.
\end{acknowledgements}

\end{document}